\documentclass[english]{IEEEtran}
\usepackage[english]{babel}
\usepackage[nocompress]{cite}
\usepackage{amsmath,amssymb,amsfonts,float}
\usepackage{algorithmic}
\usepackage{algorithm}
\usepackage{graphicx}
\usepackage{textcomp}
\usepackage{xcolor}
\usepackage{url}
\usepackage{algorithm,algorithmic}

\begin{document}
\title{Iterative Interference Cancellation for Near-Field XL-MIMO Networks with LoS Channels}

\author{Nayyab Haider and Rodrigo C. de Lamare
\thanks{N. Haider and R. C. de Lamare are with the Pontifical 
Catholic University of Rio de Janeiro (PUC-Rio), Rio de Janeiro, 
Brazil (e-mails: nayyabhaider77@hotmail.com; delamare@puc-rio.br).}%
}

\maketitle
%\markboth{XLIV BRAZILIAN SYMPOSIUM ON TELECOMMUNICATIONS AND 
%SIGNAL PROCESSING - SBrT 2026, SEPTEMBER 29TH TO OCTOBER 2ND, 
%2026, SALVADOR, BA}{}

\begin{abstract}
In extra-large multiple-input multiple-output (XL-MIMO) systems,
near-field users experience spherical wavefront propagation, offering
higher spatial resolution than far-field plane-wave models, but
complicating signal detection and interference cancellation. Under
line-of-sight conditions, we propose a weighted-rate successive
interference cancellation (WR-SIC) scheme whose ordering metric exploits
the enhanced spatial degrees of freedom of near-field propagation to
prioritize users with stronger channels. An iterative detection and
decoding (IDD) scheme with LDPC coding is then developed. Simulations in
hybrid near- and far-field deployments show that the proposed WR-SIC and
IDD schemes improve bit-error and sum rates over existing schemes.

\end{abstract}

\section{Introduction}
Recent work on extra-large multiple-input multiple-output (XL-MIMO) systems \cite{jvece,snsxl,losxl}, where massive antenna arrays with apertures with several meters enable unprecedented spatial multiplexing capabilities beyond massive multiple-input multiple-output (MIMO) systems \cite{mmimo,wence}, has challenged traditional far-field assumptions and attracted a great deal of attention. In particular, when users operate within the Rayleigh distance, electromagnetic waves propagate as spherical rather than plane wavefronts  \cite{10220205, 9903389}. This near-field behavior creates spatial non-stationarity and visibility regions while offering enhanced spatial resolution and interference management capabilities. The coexistence of near-field and far-field users in practical deployments creates hybrid scenarios that present challenges for signal detection and interference cancellation \cite{10185619}.

In this context, successive interference cancellation (SIC) is fundamental for multiple-antenna systems, with existing ordering strategies including channel norm-based \cite{cho2010mimo,7763844} and SINR-based approaches \cite{1599603,5534921}. The V-BLAST architecture with zero-forcing SIC has been extensively analyzed \cite{1378868}, while recent work extends SIC to multiuser detection with MMSE processing \cite{7636865,8842666}. Furthermore, iterative detection and decoding (IDD) has shown significant gains when combined with SIC \cite{8842666, 7105928, 1013125}, while near-field channel modeling has gained attention due to unique Fresnel region characteristics \cite{9693928,8949454}. Recent advances address spatial non-stationarity and visibility regions \cite{8638522,10509715, 9940939}, parametric channel estimation \cite{10476971}, and holographic MIMO communications \cite{10558821}. Despite the progress in interference cancellation, IDD techniques, and near-field modeling, existing literature lacks integrated frameworks that exploit near-field spatial multiplexing characteristics. Comprehensive surveys on near-field XL-MIMO communications \cite{10496996} highlight the need for novel signal processing approaches that fully exploit the enhanced spatial degrees of freedom available in these systems. 

In this work, we investigate iterative interference cancellation for extra-large multiple-input multiple-output (XL-MIMO) systems in near-field scenarios under line-of-sight (LoS) conditions. We develop an MMSE-based successive interference cancellation scheme with weighted-rate ordering, and integrate it with an iterative detection and decoding (IDD) scheme employing low-density parity-check (LDPC) codes. Our contributions are: (1) a weighted-rate ordering 
metric exploiting near-field rank enhancement; (2) integration 
with IDD for XL-MIMO; (3) performance analysis in hybrid 
near/far-field deployments with sum-rate and complexity analysis.

This work is organized as follows. Section II presents the system model. Section III introduces the SIC MMSE receiver with weighted-rate ordering and a sum-rate analysis, whereas Section IV details the proposed IDD scheme. Section V presents the results, while Section VI draws the conclusions. \vspace{-0.25em}

\section{System Model}

We consider the uplink of a near-field XL-MIMO system in which a base station (BS) equipped with $M$ receive antennas serves $K$ users, each equipped with $N_U$ transmit antennas. The system operates at a carrier frequency of $f_c = 7.5$ GHz with a corresponding wavelength $\lambda = c/f_c$, where $c$ is the speed of light. The BS employs a uniform linear array (ULA) with antenna spacing of $d = \lambda/2$. Each user transmits independent data streams from their antennas with equal power $P_k/N_U$ per antenna (total power $P_k$), resulting in a total of $K \times N_U$ spatial streams that must be detected at the BS. No channel state information feedback to users is required: only the BS needs channel knowledge for detection and ordering.

This work focuses on LoS-dominant scenarios typical in near-field XL-MIMO deployments. Extension to multipath environments is left for future work. The electromagnetic radiation field surrounding antenna arrays can be divided into two distinct regions based on the propagation characteristics: the near-field region and the far-field region. The boundary between these regions is given by the Rayleigh distance, which depends on the array geometry and the operating frequency:
\begin{equation}
R = \frac{2D^2}{\lambda}
\end{equation}
where $D$ represents the maximum array aperture dimension. Users located at distances $d < R$ from the BS operate in the near-field region and experience spherical wavefront propagation, while users at distances $d \geq R$ are in the far-field region with plane wave characteristics. \vspace{-0.5em}

\subsection{Near-Field Channel Modeling} 

For users operating in the near-field LoS region, electromagnetic waves propagate as spherical wavefronts rather than the commonly assumed plane waves. This introduces significant complexity in channel modeling but also provides enhanced spatial discrimination capabilities. The near-field channel coefficient between the $m$-th transmit antenna of user $k$ and the $n$-th receive antenna at the BS is modeled by

\begin{equation}
h_{n,m,k} = \sqrt{\beta_{n,m,k}}\; e^{-j\frac{2\pi}{\lambda} d_{n,m,k}}
\label{eq:h_nf}
\end{equation}

Here $\beta_{n,m,k}$ denotes the Friis free-space power gain \cite{1697062}.
To state the power scaling explicitly, the channel coefficient is written with
$\beta_{n,m,k}$ under a square root, so that the received power per antenna is
$|h_{n,m,k}|^2 = \beta_{n,m,k}$, following the
non-uniform spherical-wave model of \cite{10496996}.

\begin{equation}
\beta_{n,m,k} = \frac{\lambda^2}{(4\pi)^2 d_{n,m,k}^2}
\label{eq:friis}
\end{equation}
with $d_{n,m,k}$ being the Euclidean distance between user antenna and each antenna of the BS that is calculated to capture the spherical wavefront characteristics as
\begin{equation}
d_{n,m,k} = \sqrt{\sum_{i \in \{x,y,z\}} (i_{tx,k,m} - i_{rx,n})^2}
\label{eq:distance}
\end{equation}

where $i_{tx,k,m}$ and $i_{rx,n}$ represent the coordinates (x, y, or z) of the $m$-th transmit antenna of user $k$ and the $n$-th receive antenna at the BS, respectively.

This distance calculation produces spatial non-stationarity, 
nonlinear phase relationships across antenna elements, and 
distance-dependent path loss variations, all of which 
provide enhanced spatial discrimination for interference  cancellation.

\subsection{Far-Field Channel Modeling} 
In contrast to the near-field scenario, users located in the far-field LoS region can be modeled using the plane wave approximation. This simplifies the channel modeling while maintaining reasonable accuracy for users at sufficient distances. The far-field channel coefficient is expressed by

\begin{equation}
h_{n,m,k} = \sqrt{\beta_{k}}\; e^{-j\frac{2\pi}{\lambda}\,\mathbf{k}_k\cdot\mathbf{r}_n}
\label{eq:h_ff}
\end{equation}
where $\beta_k$ is a distance-independent path loss coefficient for user $k$ given by
\begin{equation}
\beta_k = \frac{\lambda^2}{(4\pi)^2 r_k^2}
\label{eq:ff_pathloss}
\end{equation}
Here, $r_k$ denotes the average distance from user $k$ to the BS, and the spatial variations across the array aperture are considered negligible. The wave vector $\mathbf{k}_k \in \mathbb{R}^{3 \times 1}$ encapsulates the directional information of the incident plane wave for user $k$ and is defined in terms of the spherical coordinate system:
\begin{equation}
\mathbf{k}_k = \begin{bmatrix}
\cos(\theta_{az,k}) \sin(\phi_{el,k}) \\
\sin(\theta_{az,k}) \sin(\phi_{el,k}) \\
\cos(\phi_{el,k})
\end{bmatrix}
\label{eq:wave_vector}
\end{equation}
where $\theta_{az,k}$ and $\phi_{el,k}$ are the azimuth and elevation angles of user $k$. The vector $\mathbf{r}_n \in \mathbb{R}^{3 \times 1}$ specifies the location of the $n$-th receive antenna relative to a reference point, typically the array center. 
In the simulations, $\theta_{az,k}$ and $\phi_{el,k}$ are
drawn independently and uniformly from $[0^\circ, 360^\circ)$
and $[0^\circ, 90^\circ]$, respectively, ensuring all users
have distinct angular orientations with probability one
\cite{10185619}. All channel parameters are assumed
perfectly known at the BS, consistent with the perfect CSI
assumption\cite{9693928}.
\vspace{-1em}

\subsection{Hybrid Network Configuration and Signal Model} 

We consider a hybrid network with $K_{NF}$ near-field and $K_{FF}$ far-field users, where $K_{NF} + K_{FF} = K$, applying spherical and plane wave models, respectively.

The channel matrix is constructed by concatenating channel vectors from all users:
\begin{equation}
\mathbf{H} = \begin{bmatrix}
\mathbf{h}_{1,1} & \mathbf{h}_{1,2} & \cdots & \mathbf{h}_{1,N_U} & \mathbf{h}_{2,1} & \cdots & \mathbf{h}_{K,N_U}
\end{bmatrix}
\end{equation}
where $\mathbf{h}_{k,n_u} \in \mathbb{C}^{M \times 1}$ represents the channel response from the $n_u$-th antenna of user $k$ to all $M$ BS antennas:
\begin{equation}
\mathbf{h}_{k,n_u} = [h_{1,n_u,k}, h_{2,n_u,k}, h_{3,n_u,k}, \ldots, h_{M,n_u,k}]^T
\end{equation}
The $M \times K N_U$ channel matrix $\mathbf{H}$ for $M$ receive antennas and $K N_U$ spatial streams can be partitioned into user-specific sub-matrices as given by
\begin{equation}
\mathbf{H} = [\mathbf{H}_1, \mathbf{H}_2, \mathbf{H}_3, \ldots, \mathbf{H}_K]
\end{equation}
where each $\mathbf{H}_k \in \mathbb{C}^{M \times N_U}$ contains responses from all antennas of user $k$ as described by $\mathbf{H}_k = [\mathbf{h}_{k,1}, \mathbf{h}_{k,2}, \ldots, \mathbf{h}_{k,N_U}]$.

The received signal represents the superposition of all transmitted streams with noise and is given by
\begin{equation}
\begin{split}
\mathbf{y} &= \sum_{k=1}^{K_{NF}} \sum_{n_u=1}^{N_U} \mathbf{h}_{k,n_u} x_{k,n_u} + \sum_{k=K_{NF}+1}^{K} \sum_{n_u=1}^{N_U} \mathbf{h}_{k,n_u} x_{k,n_u} + \mathbf{n}\\
&= \sum_{k=1}^{K_{NF}} \mathbf{H}_k \mathbf{x}_k + \sum_{k=K_{NF}+1}^{K} \mathbf{H}_k \mathbf{x}_k + \mathbf{n}
\end{split}
\end{equation} \vspace{-0.1em}
where $\mathbf{y} \in \mathbb{C}^{M \times 1}$ is the received signal vector, $\mathbf{H} \in \mathbb{C}^{M \times K \cdot N_U}$ is the channel matrix, $\mathbf{x} \in \mathbb{C}^{K \cdot N_U \times 1}$ is the transmitted symbol vector, and $\mathbf{n} \sim \mathcal{CN}(0, \sigma_n^2 \mathbf{I}_{M})$ is the noise vector. \vspace{-1em}

\section{Proposed SIC MMSE Receiver with Weighted-Rate (WR) Ordering}
In this section, we present the proposed SIC MMSE receiver with
weighted-rate ordering that exploits the unique spatial
characteristics of near-field XL-MIMO systems. We develop the mathematical framework for the weighted-rate metric computation, present the ordering algorithm, and analyze the interference cancellation process along with computational complexity.

\subsection{Weighted-Rate Metric and MMSE Filter}

In near-field XL-MIMO systems, the per-element distance
computation in \eqref{eq:distance} produces nonlinear phase
profiles across the array for each user. By contrast, the
far-field model of \eqref{eq:h_ff}--\eqref{eq:wave_vector}
imposes a linear phase gradient, yielding a rank-1 or low-rank
channel structure per user \cite{9903389}. This nonlinear
phase diversity leads to richer singular value distributions
and higher effective channel rank \cite{10220205,
9903389}, as explicitly demonstrated for ULA
configurations in \cite{10496996}.

We first perform a singular value decomposition (SVD) of the channel matrix:
\begin{equation}
\mathbf{H} = \mathbf{U} \boldsymbol{\Sigma} \mathbf{V}^*
\label{eq:svd_decomposition}
\end{equation}
where $\boldsymbol{\Sigma} = \text{diag}(\sigma_1, \sigma_2, \ldots, \sigma_{\min(M, K \cdot N_U)})$ with $\sigma_1 \geq \sigma_2 \geq \ldots \geq \sigma_{\min(M, K \cdot N_U)}$.

The total spatial degrees of freedom are given by
\begin{equation}
\text{DoF} = \text{rank}(\mathbf{H}) \leq \min(M, K \cdot N_U)
\label{eq:dof}
\end{equation}

Under the considered system setup, let $\mathcal{P}$ denote
the uniform spatial distribution of user positions within the
near-field region ($d < R$), modeled via
\eqref{eq:h_nf}--\eqref{eq:distance}, and at equivalent
distances under the plane wave approximation of
\eqref{eq:h_ff}--\eqref{eq:wave_vector}. Then, the
expected effective rank of the channel matrix $\mathbf{H}$
satisfies:
\begin{equation}
\mathbb{E}_{\mathcal{P}}[\text{rank}(\mathbf{H})|_{\text{NF}}]
\geq
\mathbb{E}_{\mathcal{P}}[\text{rank}(\mathbf{H})|_{\text{FF}}]
\label{eq:rank_comparison}
\end{equation}
where $\mathbb{E}_{\mathcal{P}}[\cdot]$ denotes expectation
over $\mathcal{P}$, and the subscripts NF and FF indicate
evaluation under the near-field and far-field channel models,
respectively. This rank increase is due to the distance-dependent quadratic phase term
$\pi n^2 d^2 \cos^2\theta_{k,m}/(\lambda r_{k,m})$ in the near-field array
response, which vanishes in the far field. Two users at the same angle but
different distances have linearly independent near-field channel vectors,
whereas their far-field channel vectors are collinear. This distance-domain
separability raises the effective rank over the user distribution
\cite{10220205, 10496996}.

The weighted-rate metric for the $k$-th user exploits the singular value structure from \eqref{eq:svd_decomposition} and is defined by
\begin{equation}
M_k = \sum_{i=1}^{N_U} \sigma_{k,i}^2 \log_2 \left(1 + \frac{P_k \sigma_{k,i}^2}{\sigma_n^2} \right),
\label{eq:multiplexing_gain}
\end{equation}
where $\sigma_{k,i}$ are the singular values of the user-specific channel matrix $\mathbf{H}_k$ obtained from:
\begin{equation}
\mathbf{H}_k = \mathbf{U}_k \boldsymbol{\Sigma}_k \mathbf{V}_k^*
\label{eq:user_svd}
\end{equation}
with $\boldsymbol{\Sigma}_k = \text{diag}(\sigma_{k,1}, \sigma_{k,2}, \ldots, \sigma_{k,N_U})$.

The optimal linear MMSE receive filter is well-known to be \cite{cho2010mimo}:
\begin{equation}
\mathbf{W}_{\text{MMSE}} = \mathbf{H}^H(\mathbf{H}\mathbf{H}^H + \sigma_n^2\mathbf{I}_M)^{-1}
\label{eq:mmse_filter}
\end{equation}

\subsection{Proposed Ordering and Interference Cancellation}
The proposed ordering algorithm prioritizes users with the highest weighted-rate metric from \eqref{eq:multiplexing_gain}. The metric values are computed for all users using their individual SVD from \eqref{eq:user_svd} and sorted to obtain the detection order as follows:
\begin{equation}
M_{k_1} \geq M_{k_2} \geq \ldots \geq M_{k_K}, 
\label{eq:sorting}
\end{equation}
yielding the detection order $\pi = [k_1, k_2, \ldots, k_K]$.

The proposed SIC MMSE receiver with weighted-rate ordering involves the following steps.
In the initialization, we set the received vector ${\mathbf y}$ as the vector for the first SIC stage and the set of $K \cdot N_U$ data streams given by
\begin{equation}
\mathbf{y}^{(1)} = \mathbf{y}, \quad \mathcal{R}^{(1)} = \{1, 2, \ldots, K \cdot N_U\}
\label{eq:initialization}
\end{equation}

For the $K \cdot N_U$ data streams, we perform user selection based on the weighted-rate ordering in \eqref{eq:sorting}:
\begin{equation}
k_i^* = \pi(i)
\label{eq:user_selection}
\end{equation}

For each user $k_i^*$, individual streams are ordered by descending singular values $\sigma_{k_i^*,1} \geq \sigma_{k_i^*,2} \geq \ldots \geq \sigma_{k_i^*,N_U}$ from the user-specific SVD in \eqref{eq:user_svd}.

The MMSE receive filter for this stream is applied to the residual received
signal $\mathbf{y}^{(i)}$ to obtain the soft estimate:
\begin{equation}
\bar{x}_{k_i^*} = \mathbf{w}_{k_i^*}^H \mathbf{y}^{(i)}
\label{eq:symbol_detection}
\end{equation}
where $\mathbf{w}_{k_i^*}$ is the corresponding row of $\mathbf{W}_{\text{MMSE}}$
in \eqref{eq:mmse_filter}. The soft estimate is then mapped to the nearest
constellation point by a symbol-by-symbol minimum-distance decision:
\begin{equation}
\hat{x}_{k_i^*} = \mathcal{Q}\!\left(\bar{x}_{k_i^*}\right)
\label{eq:hard_decision}
\end{equation}
where $\mathcal{Q}(\cdot)$ denotes the hard-decision quantizer that maps the
MMSE soft estimate to the nearest constellation symbol. The contribution of
the detected symbol is then subtracted from the residual:
\begin{equation}
\mathbf{y}^{(i+1)} = \mathbf{y}^{(i)} - \mathbf{h}_{k_i^*} \hat{x}_{k_i^*}
\label{eq:interference_cancellation}
\end{equation}

In the last step of each SIC stage, we remove the detected stream from the remaining set of data streams:

\begin{equation}
\mathcal{R}^{(i+1)} = \mathcal{R}^{(i)} \setminus \{i\}
\label{eq:update}
\end{equation}

Unlike channel-norm ordering, which is rate-blind, and capacity-based
ordering, which weights all streams equally, the proposed metric weights
each stream's rate $\log_2(1 + P_k\sigma_{k,i}^2/\sigma_n^2)$ by its squared
singular value $\sigma_{k,i}^2$, prioritizing streams that are both reliable
to cancel first and information-rich, which limits error propagation in SIC
\cite{10220205, 9903389,rsbd}. The computational cost of the proposed and existing linear and SIC receivers is shown in TABLE I.

\begin{table}[htb]
\centering
\caption{Computational Cost Comparison}
\label{tab:complexity_comparison}
\small
\begin{tabular}{|l|c|}
\hline
\textbf{Detection Scheme} & \textbf{Complexity} \\
\hline
Linear MMSE & 
$\mathcal{O}(\min(M^3, (KN_U)^3))$ \\
\hline
Channel Norm SIC \cite{cho2010mimo} & 
$\mathcal{O}(K^2 \cdot M \cdot N_U)$ \\
\hline
SINR-based SIC \cite{1599603} & 
$\mathcal{O}(K^2 \cdot M \cdot N_U)$ \\
\hline
Proposed WR-SIC & 
$\mathcal{O}(M^3 + K \cdot M \cdot N_U^2)$ \\
\hline
\end{tabular}
\end{table}

\subsection{Sum-Rate Analysis}

The sum-rate performance of the proposed weighted-rate SIC receiver can be obtained by assuming Gaussian signaling and considering the effective SINR after each stage.
Following the standard turbo-IDD analysis methodology
\cite{7105928, 1013125}, the Gaussian
signaling assumption provides a theoretical upper bound on the
achievable rate per stream, while the simulations in
Section~V employ QPSK with LDPC coding operating below
this bound.

For the $i$-th detected stream, the achievable rate is given by
\begin{equation}
R_i = \log_2\left(1 + \text{SINR}_i^{\text{eff}}\right)
\label{eq:rate_stream}
\end{equation}
The effective SINR for the $i$-th stream after SIC is:
\begin{equation}
\text{SINR}_i^{\text{eff}} = \frac{P_{k_i^*} |\mathbf{w}_{k_i^*}^H \mathbf{h}_{k_i^*}|^2}{\sum_{j \in \mathcal{R}^{(i+1)}} P_j |\mathbf{w}_{k_i^*}^H \mathbf{h}_j|^2 + \sigma_n^2 \|\mathbf{w}_{k_i^*}\|^2}
\label{eq:sinr_effective}
\end{equation}
where $\mathbf{w}_{k_i^*}$ is the MMSE filter for the $k_i^*$-th user from \eqref{eq:mmse_filter}, and $\mathcal{R}^{(i+1)}$ is the remaining stream set from \eqref{eq:update}.
The total achievable sum-rate for the system is:
\begin{equation}
R_{\text{sum}} = \sum_{i=1}^{K \cdot N_U} R_i = \sum_{i=1}^{K \cdot N_U} \log_2\left(1 + \text{SINR}_i^{\text{eff}}\right)
\label{eq:sum_rate}
\end{equation}
For near-field scenarios, the enhanced spatial degrees of freedom from \eqref{eq:rank_comparison} lead to improved sum-rate performance:
\begin{equation}
\mathbb{E}_{\mathcal{P}}[R_{\text{sum}}^{NF}] \geq
\mathbb{E}_{\mathcal{P}}[R_{\text{sum}}^{FF}]
\label{eq:sum_rate_comparison}
\end{equation}
This is achieved by the weighted-rate ordering from \eqref{eq:multiplexing_gain}, which prioritizes users with better channel conditions, reducing error propagation and maximizing system throughput.

%\subsection{Near-Field Specific Advantages}

%The proposed capacity-weighted metric exploits near-field characteristics through several mechanisms. First, the $\sigma_{k,i}^2$ weighting in \eqref{eq:multiplexing_gain} naturally favors near-field users whose channels exhibit stronger gains due to proximity. According to \eqref{eq:friis}, the path loss coefficient $\beta_{n,m,k} \propto 1/d_{n,m,k}^2$ creates distance-dependent weighting that benefits closer users. Second, unlike far-field plane waves where phase variations are linear (Eq. 7), near-field spherical wavefronts create nonlinear phase relationships (Eq. 2) that result in richer singular value distributions. Third, spatial non-stationarity is explicitly exploited where different antenna elements experience significantly different channel conditions (varying $d_{n,m,k}$), creating diverse singular values that the metric weighs in the ordering decision. While the metric could theoretically apply to any high-rank channel, it is particularly effective in near-field LoS scenarios where the rank enhancement arises from geometric properties of spherical wavefront propagation rather than scattering.

\section{Proposed Iterative Detection and Decoding}

The proposed IDD scheme integrates the WR-SIC with LDPC coding through iterative soft information exchange \cite{7105928, 1013125,spa,mfsic,mbdf,idd1bit,cdidd,msgamp,llraps,comp,iddocl,icce,nfidd}, but differs from \cite{7105928} (which 
addresses 2×2 block-fading MIMO with no ordering) by 
operating on near-field XL-MIMO LoS channels with 
spherical wavefront geometry \eqref{eq:h_nf}--\eqref{eq:distance}, incorporating $M_k$ 
from \eqref{eq:multiplexing_gain} in each iteration, and targeting hybrid 
near/far-field deployments. A rate $1/2$ LDPC code with 
log-domain belief propagation is employed, exchanging 
Log-Likelihood Ratios (LLRs) between detector and decoder until convergence.

\subsection{Scalar Equivalent Model and LLR Computation}

Each data stream is modeled using a scalar Gaussian approximation:
\begin{align}
\bar{x}_{\pi(i)} &= \mu_{\pi(i)} x_{\pi(i)} + \xi_{\pi(i)} \label{eq:detection} \\
\tilde{x}_{\pi(i)} &= \text{LDPC-Decoder}(\text{LLR}(\bar{x}_{\pi(i)})) \label{eq:postdecoding}
\end{align}
where $\bar{x}_{\pi(i)}$ represents the output of the receive filter before LDPC decoding, $\mu_{\pi(i)} = \mathbf{w}_{\pi(i)}^H \mathbf{h}_{\pi(i)}$ is the effective gain, and $\xi_{\pi(i)} \sim \mathcal{CN}(0, \sigma^2_{\pi(i)})$ represents residual interference plus noise. The refined estimates $\tilde{x}_{\pi(i)}$ obtained from the LDPC decoder output are used for interference cancellation in subsequent IDD iterations.

The log-likelihood ratio (LLR) for bit $b_l$ of symbol $x_k$ is:
\begin{equation}
L_D(b_l) = \log \frac{\sum_{x \in \mathcal{X}^{+1}_l} \exp\left(-\frac{|\bar{x}_k - \mu_k x|^2}{\sigma^2_k}\right) P(x)}{\sum_{x \in \mathcal{X}^{-1}_l} \exp\left(-\frac{|\bar{x}_k - \mu_k x|^2}{\sigma^2_k}\right) P(x)} - L_C(b_l)
\end{equation}
where $\mathcal{X}^{+1}_l$ and $\mathcal{X}^{-1}_l$ are constellation symbol sets with bit $b_l = +1$ and $b_l = -1$, and $L_C(b_l)$ is the extrinsic LLR from the LDPC decoder.

The a priori probability of symbol $x$ is:
\begin{equation}
P(x) = \prod_{m=1}^{M_c} \frac{1}{1 + \exp(-x^b_m L_C(b_m))}
\end{equation}
where $M_c$ is the number of bits per symbol.

The residual variance after SIC processing is:
\begin{equation}
\sigma^2_k = \sigma^2_n \|\mathbf{w}_k\|^2 + \sum_{j \neq k} \text{Var}(x_j) |\mathbf{w}_k^H \mathbf{h}_j|^2
\end{equation}

\subsection{Message Passing Algorithm}

For each iteration $t$, the detector computes extrinsic LLRs:
\begin{equation}
L_D^{(t)}(b_l) = \log \frac{\sum_{x \in \mathcal{X}^{+1}_l} \exp\left(-\frac{|\bar{x}_k^{(t)} - \mu_k^{(t)} x|^2}{\sigma^2_{k,t}}\right) P^{(t-1)}(x)}{\sum_{x \in \mathcal{X}^{-1}_l} \exp\left(-\frac{|\bar{x}_k^{(t)} - \mu_k^{(t)} x|^2}{\sigma^2_{k,t}}\right) P^{(t-1)}(x)}
\end{equation}
The LDPC decoder updates a priori probabilities as:
\begin{equation}
P^{(t)}(x) = \prod_{m=1}^{M_c} \frac{\exp(x^b_m L_C^{(t)}(b_m)/2)}{2\cosh(L_C^{(t)}(b_m)/2)}
\end{equation}
The process continues until the convergence of the LDPC decoder, the maximum number of iterations, or the LLR threshold is reached.

\section{Numerical Results}
\label{section_V}
The evaluation of the proposed and existing interference cancellation schemes is conducted through Monte Carlo simulations. The simulation parameters are carefully selected to represent realistic near-field XL-MIMO scenarios while enabling direct comparison between different techniques. The LDPC codes are constructed with the PEG algorithm \cite{peg,memd}. We assume that the interference cancellation schemes have perfect channel state information. Table \ref{tab:simulation_params} gives the simulation parameters.

\begin{table}[h!]
\centering
\caption{Simulation Parameters}
\label{tab:simulation_params}
\begin{tabular}{|l|c|}
\hline
\textbf{Parameter} & \textbf{Value} \\
\hline
Number of BS receive antennas ($M$) & 64 \\
\hline
Number of users ($K$) & 16 \\
\hline
Transmit antennas per user ($N_U$) & 2 \\
\hline
Total spatial streams ($K \times N_U$) & 32 \\
\hline
Antenna spacing ($d$) & $\lambda/2$ \\
\hline
Modulation scheme & QPSK \\
\hline
Channel coding & LDPC (rate 1/2) \\
\hline
Maximum IDD iterations & 3 \\
\hline
LDPC decoder iterations & 20
\\
\hline
\end{tabular}
\end{table}

For the hybrid scenario, half the users are randomly placed in
the near-field region ($d < R$) and half in the far-field
($d \geq R$), with distances drawn uniformly from
$[1.2D,\,R)$ and $[R,\,3R]$, respectively \cite{10185619}.
 Each simulation point averages at least $10^4$ independent
channel realizations. For low BER points, additional codewords
are transmitted over the same $10^4$ realizations to reduce
variance.

\begin{figure}[h!]
\centering
    \includegraphics[width=0.4\textwidth]{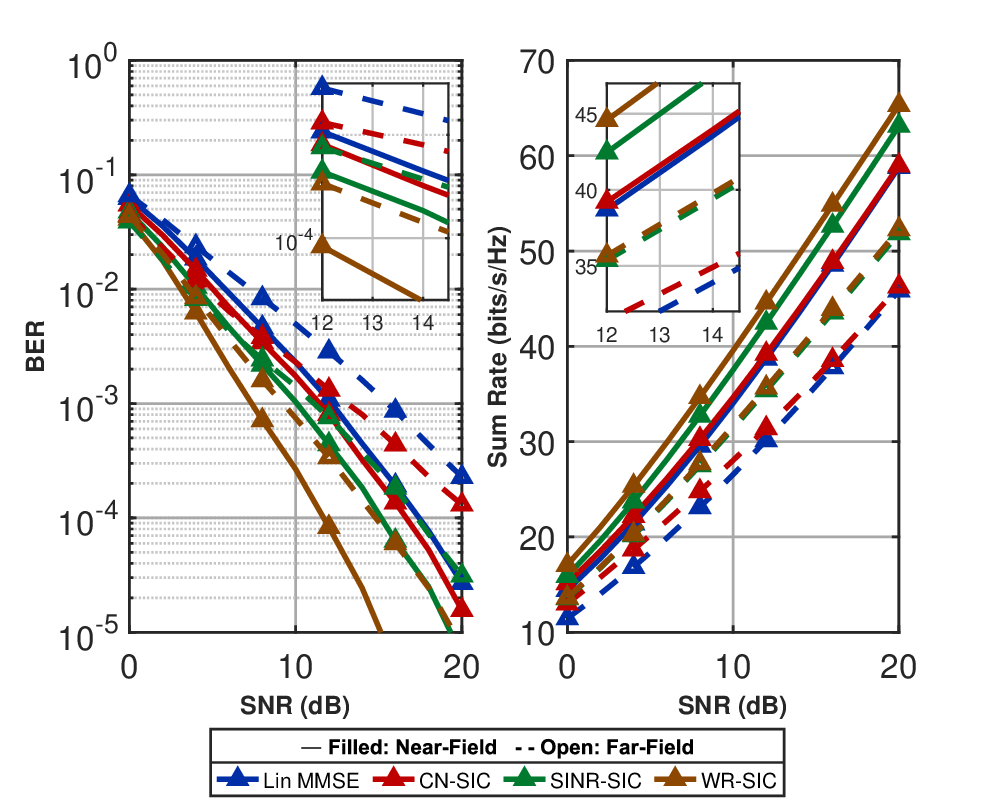}
    %fig_conf1.eps
 \vspace{-0.5em}
\caption{\small BER and sum rate performance 
comparison of near and far field users under perfect CSI}
\label{fig: BER SR perfect csi}
\end{figure}

The results show superior performance of the proposed WR-SIC
over conventional detection schemes across all scenarios.
The left plot of Fig. \ref{fig: BER SR perfect csi} shows
the average BER performance comparison between near-field
and far-field users for all detection techniques. Near-field users (solid) outperform far-field users (dashed), 
with the gap increasing at higher SNR due to spherical 
wavefront spatial resolvability. The proposed
WR-SIC achieves approximately 2--3~dB SNR gain compared to
channel norm \cite{cho2010mimo} and SINR-based SIC
\cite{1599603} at BER of $10^{-3}$, with linear MMSE
exhibiting the poorest performance across both user categories.

The right plot of Fig. \ref{fig: BER SR perfect csi}
presents the sum-rate performance comparison between
near-field and far-field users using different detection
schemes. The results corroborate the BER findings, showing
that near-field users consistently achieve higher sum-rates
than their far-field counterparts. The proposed WR-SIC
demonstrates the highest sum-rate performance for both types
of users, exploiting the enhanced spatial discrimination
capabilities of near-field propagation. 
%\begin{figure}[h!]
%\centering
%\begin{subfigure}[h!]{0.75\columnwidth}
    %\includegraphics[width=\textwidth]{figure_2.pdf}
    %\caption{Average BER performance}
    %\label{fig:ber_nf_ff}
%\end{subfigure}
%\hfill
%\begin{subfigure}[h!]{0.75\columnwidth}
    %\includegraphics[width=\textwidth]{figure_5.pdf}
    %\caption{Sum rate performance}
    %\label{fig:sumrate_nf_ff}
%\end{subfigure}
%\vspace{-0.5em}
%\caption{Performance comparison for near and far-field users}
%\label{fig:performance_comparison}
%\end{figure}
 \vspace{-1em}

%\begin{figure}[h!]
%\centering
%\includegraphics[width=0.75\columnwidth]{idd_final.pdf}
%\vspace{-0.5em}
%\caption{Average BER over all $K=16$ users (half near-field,
%half far-field) with and without IDD using 1 and 3 iterations} 
%\label{fig:ber_idd}
%\end{figure}
\begin{figure}[h!]
\centering
    \includegraphics[width=0.4\textwidth]{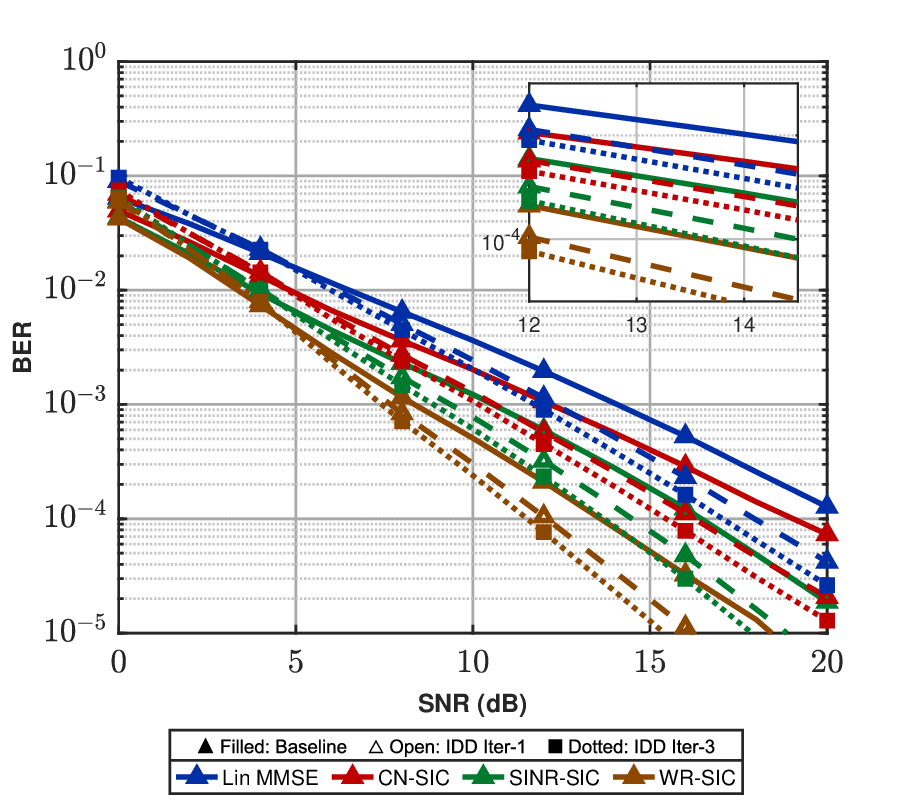}
    %conffig2.eps
 \vspace{-0.5em}
\caption{Average BER over all users (half near-field, half far-field) with and without IDD using 1 and 3 iterations} 
\label{fig: IDD_perfect csi}
\end{figure} \vspace{-0.1em}

Fig.~\ref{fig: IDD_perfect csi} reports the average BER over all $K=16$ users,
providing a system-level view of IDD gains across the
full hybrid user population. All detection schemes benefit from iterative
processing, with the proposed WR-SIC achieving approximately
1--2~dB additional SNR gain at BER of $10^{-3}$ with just
1 iteration. Extending to 3 IDD iterations provides further
enhancements across all methods. The diminishing returns with
extra iterations suggest that most gains are captured in the
first iterations, making 3 iterations practical.

\section{Conclusion}
%XL-MIMO systems call for novel signal processing approaches that exploit near-field propagation characteristics. 
This work presented an interference cancellation strategy combining a novel weighted-rate SIC ordering technique and an IDD scheme for LoS channels. Unlike conventional channel norm or SINR-based ordering, the proposed method exploits enhanced spatial degrees of freedom through weighted-rate analysis that leverages the geometric properties of spherical wavefront propagation. The results show superior performance of near-field users to far-field users, validating the effectiveness of the proposed approach in XL-MIMO systems. \vspace{-0.95em}

\bibliographystyle{IEEEtran}
\bibliography{references3}

\end{document}